\documentclass{PoS}

\def\gsim{\mathrel{\raise.5ex\hbox{$>$}\mkern-14mu
             \lower0.6ex\hbox{$\sim$}}}
\def\lsim{\mathrel{\raise.3ex\hbox{$<$}\mkern-14mu
             \lower0.6ex\hbox{$\sim$}}}

\newcommand{\aap}{    {\it Astron. Astrophys.}}

\newcommand{\apj}{    {\it Astrophys. J.}}
\newcommand{\apjl}{    {\it Astrophys. J. Lett.}}

\newcommand{\mnras}{  {\it Mon. Not. Roy. Astron. Soc.}}

\newcommand{\ssr}{    {\it Space Sci. Rev.}}

\title{A Cosmic Battery in accretion flows around astrophysical black holes}

\ShortTitle{A Cosmic Battery}

\author{\speaker{Ioannis Contopoulos}\thanks{In memory of father Constantinos Stratigopoulos.}
\\
        Research Center for Astronomy and Applied Mathematics, Academy of Athens\\
        E-mail: \email{icontop@academyofathens.gr}}

\abstract{We present our view of the immediate environment of an astrophysical black hole. We discuss the origin of the electron-positron and electron-proton jet components, and the origin of the large-scale magnetic field. We show how the aberrated radiation field viewed by the rotating plasma generates an azimuthal induction electric field, and how the curl of this electric field steadily grows the magnetic field to equipartition over a few hundred million dynamical times. That timescale corresponds to weeks for stellar mass black holes, and million years for super-massive black holes. We conclude with numerical and observational confirmation of the action of a Cosmic Battery in accretion flows around astrophysical black holes.}

\FullConference{International Conference on Black Holes as Cosmic Batteries: UHECRs and Multimessenger Astronomy - BHCB2018\\
		12-15 September, 2018\\
		Foz du Iguazu, Brasil}

\begin{document}

\section{The immediate environment of an astrophysical black hole}

The primary goal of the Event Horizon Telescope (EHT) has been to image the black hole with the largest apparent event horizon, namely the four million solar mass black hole at the center of our Galaxy. After about two years of observations and analyses, the EHT may not be able to achieve its primary goal. We expect, however, that it will teach us a few things about the immediate environment of an astrophysical black hole, and in particular what is the structure of the innermost relativistic jet, the origin of the matter in the jet, the structure of the innermost magnetic field, etc.

Another future target of the EHT is the center of the M87 galaxy in the Virgo cluster. We have been greatly impressed by observations of a spine component in the M87 radio jet by \cite{ANP16} and \cite{Metal16}. Currently, such observations can only be obtained through space VLBI with the VSOP satellite. It seems that the jet consists of three sub-components: a very narrow, collimated, and difficult to observe one that we interpret as the Blandford-Znajek electron-positron jet (the `spine' \cite{BZ77}), an outer parabolic one that we interpret as the innermost edge of the electron-proton disk wind (the `sheath'), and an in-between region of weaker radio emission. The sheath transitions from parabolic to conical beyond a distance of about 100~pc that we associate with the Bondi radius for M87 \cite{AN12}. It is interesting that transverse gradients of the Faraday rotation measure (hereafter RM) have been observed over several parts of the large scale conical radio jet \cite{AAN13}. Such gradients have been associated with the jet itself \cite{Cetal16}, and prove that the sheath consists of electrons and protons, since an electron-positron plasma does not produce Faraday rotation. This may either confirm that the sheath indeed originates in the accretion disk around the black hole, or that the Blandford-Znajek electron-positron jet is mass-loaded (`contaminated') with ordinary plasma from the environment through which it travels \cite{PBB17}. We prefer the former interpretation.

The radio jet originates somewhere in the innermost black hole magnetosphere which may after all be unobservable. Nevertheless, we can learn a lot from the analogy with pulsars. The black hole magnetosphere is a generalization of the pulsar magnetosphere, and is in some respects more complex (it has two light cylinders \cite{NC14}), and in others simpler (it does not include the region of closed magnetic field lines that made the resolution of the pulsar magnetosphere problem so hard to obtain \cite{CKF99}). There are two unknown functions in a steady-state black hole magnetosphere: the distribution of  electric charge (or equivalently the angular velocity of rotation of magnetic field lines\footnote{Black holes do not have a solid surface on to which magnetic field lines are anchored as do pulsars.}) and the distribution of electric current. A central result of the analogy with pulsars is that these two unknown functions are specified as {\em eigenfunctions} by the requirement that magnetic field lines cross the two light cylinders smoothly. In other words, we are not free to specifiy the distribution of electric charge and electric current in the black hole magnetosphere. These are obtained by the self-consistent solution of the steady-state problem. And in all cases, the solution contains {\em a null surface} above the horizon (i.e. a region of reversal of the sign of the magnetospheric electric charge \cite{LS17}), and {\em a charged current sheet} along the last open magnetic surface that threads the black hole horizon \cite{NC14}. Both regions are astrophysically very interesting. The former may be related to the origin of the electron-positron pairs in the black hole jet, while the latter may lead to particle acceleration and consequent high-energy radiation along the edge of the jet.

\section{The origin of the magnetic field}

\footnote{This section is adapted from \cite{C19}.}
All these interesting astrophysical processes require the presence of a large scale magnetic field that threads the black hole horizon and the accretion flow around it. Unfortunately, the origin of this field remains an open question. \cite{McKTB12} argued that ``observations (around AGNs) do show patches of coherent magnetic flux surrounding astrophysical systems that can feed black holes''. They reached the conclusion that even if only about 10 per cent of it manages to accrete via the accretion disk, it will be enough to generate the jets and winds that we are observing. Nevertheless, there is strong debate in the astrophysics community whether accretion can indeed bring in magnetic field from such `reservoirs' of magnetic flux located far from the central black hole \cite{vB89, LPP94, LRN94}. 

We prefer another possibility, namely that the magnetic field that threads the black hole horizon is generated by the Poynting-Robertson drag effect on the plasma electrons in a so-called `Cosmic Battery' around the central black hole \cite{CK98, C15}. In order to understand the Cosmic Battery, it is necessary to understand how radiation acts on a plasma. Radiation enters dynamically as an extra term in the equation of motion of the plasma, namely
\begin{equation}\label{eqmotionplasma}
\frac{{\rm d}{\bf v}}{{\rm d}t}=\dots + \frac{{\bf f}_{\rm rad}}{m_p}\ ,
\end{equation}
where, ${\bf v}$ is the plasma velocity, and $m_p$ is the mass of the proton (for ease of presentation, we assume here an electron-proton plasma).  It is well known that radiation acts almost exclusively on the plasma electrons, and in fact ${\bf f}_{\rm rad}$ is the radiation force per electron. So, how come radiation contributes to the dynamics of the plasma as a whole, e.g. how is it possible that radiation holds stars from collapsing under their own weight? The answer is that, in the presence of radiation, {\it an inductive electric field} ${\bf E}$ {\it develops in the interior of the plasma}. The origin of this electric field becomes clear when we consider the equations of motion both of the protons and the electrons. In the presence of radiation, the equation of motion of the protons contains an extra term
\begin{equation}\label{eqmotionprotons}
m_p\frac{{\rm d}{\bf v}_p}{{\rm d}t}=\dots + e{\bf E}\ ,
\end{equation}
and the equation of motion of the electrons also contains the radiation force term
\begin{equation}\label{eqmotionelectrons}
m_e\frac{{\rm d}{\bf v}_e}{{\rm d}t}=\dots + {\bf f}_{\rm rad} - e{\bf E}\ .
\end{equation}
Here, ${\bf v}_p, {\bf v}_e$ are the velocities of the protons and the electrons respectively, $m_e$ is the mass of the electron, and $e$ is the magnitude of the electron charge. The velocities of the electrons and the protons do not differ much and  $m_e\ll m_p$, therefore, eq.~(\ref{eqmotionelectrons}) is equal to zero to a very good approximation. Thus
\begin{equation}\label{E}
{\bf E}={\bf f}_{\rm rad}/e\ .
\end{equation}
This is how the radiation force appears in the equation of motion for the protons and the plasma as a whole. This effect is often ignored by the younger generation of researchers. Another way to understand this result is that without the electric field, the radiation force would have modified the motion of the electrons so dramatically that an enormous electric current and an associated magnetic field would have appeared in the interior of the plasma. It is not, however, possible to `turn on' an electric current inside a highly conducting plasma. Faraday's law tells us that the plasma will react, and an induction electric field will appear in the direction opposite to the direction of the growing electric current that will oppose its growth. It is thus not recommended to study the growth of the electric current and the associated magnetic field through the velocity difference between the electrons and the protons, as for example in \cite{L15}. This approach ignores the inductive reaction of the plasma, and may lead to wrong conclusions. 

The radiation force per electron must be calculated in the rest frame of the electron. Even if the radiation field is isotropic, the moving electron absorbs photons coming from the direction opposite to its direction of motion, and re-emits them isotropically. The resulting radiation force is thus equal to
\begin{equation}
{\bf f}_{\rm rad}=\frac{\sigma_{\rm T} {\bf F}_{\rm rad}}{c}
\end{equation}
where, $\sigma_{\rm T}$ is the electron Thomson cross-section, and ${\bf F}_{\rm rad}$ is the radiation flux (flow of radiation energy per unit surface) seen in the frame of the electron. The calculation of the radiation field in the vicinity of an astrophysical source of X-rays is complex and involves detailed radiation transfer with absorption, emission, and ray tracing. Another simpler not as accurate approach is to assume that the radiation field behaves as a fluid, i.e. that there exists a frame of references in which the radiation field is isotropic \cite{Cetal18}.  \cite{KC14} performed ray-tracing calculations over several optically thick and optically thin distributions of matter. One may obtain a crude esimate of the radiation force per electron in the idealized case of a central radiation source with luminosity $L$, and an electron in circular motion around it at distance $r$. In that case, 
\begin{equation}\label{PRforce}
{\bf f}_{\rm rad}=\frac{L\sigma_{\rm T}}{4\pi r^2 ce}\left(\hat{r}-\frac{v^\phi}{c}\hat{\phi}\right)\ .
\end{equation}
Notice here the analogy with the Poynting-Robertson drag force on dust grains in orbit around the Sun, only in that case the geometric cross-section of the grains is introduced \cite{P03, R37}.

We will now discuss the growth of the magnetic field generated by the intense radiation and rotational velocity fields in the innermost accretion flow around an astrophysical black hole. 
The correct way to do this is through the induction equation,
\begin{equation}\label{induction}
\frac{\partial {\bf B}}{\partial t} = -\nabla\times \left( -{\bf v}\times {\bf B} + {\bf E}c +\eta\nabla\times {\bf B}\right)\ ,
\end{equation}
where, $\eta$ is the magnetic diffusivity in the interior of the disk. Obviously, if $\nabla\times {\bf E}=0$ as in stellar interiors, radiation does not generate electric currents, neither magnetic fields. In stars, radiation pushes the electrons outwards leading to a surplus of electrons in the outer layers of the star, and a depletion in its center. Stars are electrically polarized along their radius and develop an electric potential difference between the center and the surface (about one volt in the case of the sun), but no magnetic fields are generated. As we will now see, rotation introduces electric fields with non-zero curl.

\begin{figure}[t]
\centering
\includegraphics[width=8 cm]{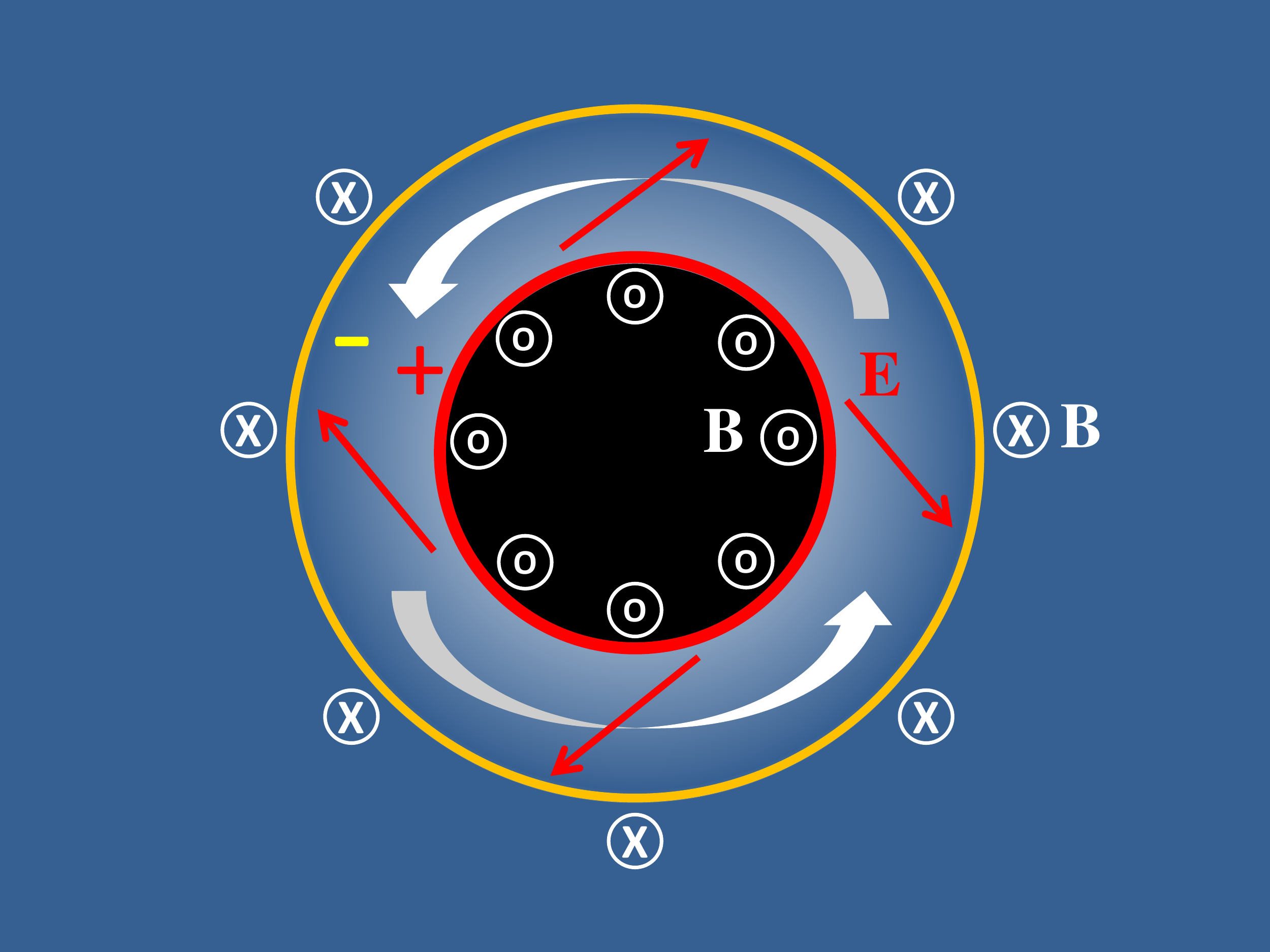}
\caption{Cartoon of the electric and magnetic fields generated by the Cosmic Battery mechanism. Rotating disk seen from above with central radiation source. Direction of rotation: counter-clockwise (white arrows). Direction of induction electric field ${\bf E}$ due to radiation: radial and clockwise (red arrows). Positive charge accumulated around the inner edge of the disk (red circle). Negative charge accumulated around the distance of optical depth one in the disk (yellow circle). Direction of generated poloidal magnetic field (small white circles): ${\bf B}$ parallel to $\Omega$ around the center, ${\bf B}$ antiparallel to $\Omega$ in the bulk of the disk.}
\label{CosmicBattery}
\end{figure}

Combining the simple expression obtained in eq.~(\ref{PRforce}), eq.~(\ref{E}), and the integrated form of eq.~(\ref{induction}), we obtain
\begin{equation}\label{inductionintegrated}
\frac{\partial \Psi}{\partial t} \approx 2\pi rc\left( ({\bf v}\times {\bf B})^\phi - \frac{f_{\rm rad}^\phi}{e} +\eta(\nabla\times {\bf B})^\phi\right)\ ,
\end{equation}
where, $\Psi$ is the magnetic flux contained inside radius $r$ (see also \cite{B-KLB02} for a generalized expression). The second term in the r.h.s. of the above equation generates a poloidal magnetic field along the direction of the angular velocity vector $\Omega$ in the disk ($f_{\rm rad}^\phi$ is negative in most parts of the disk). The magnetic flux thus generated closes in the outer parts of the disk not reached by radiation from the center where $f_{\rm rad}^\phi$ drops to zero (see cartoon in figure~\ref{CosmicBattery}). This poloidal magnetic flux will be advected inwards by the ideal accretion flow represented by the first term in eq.~(\ref{inductionintegrated}). We assume that, ideal MHD conditions apply near the central black hole, and flux accumulated inside the inner edge of the disk will keep growing. The growth will cease and the accumulated magnetic flux will saturate if the flow also carries along the return polarity of the magnetic field \cite{CK98, B-KLB02, CKC06}. However, if the return polarity lies in a region with significant magnetic diffusivity so that the third term in eq.~(\ref{inductionintegrated}) dominates over the first \cite{vB89, LPP94, LRN94} (see also, however, \cite{LRB-K09}), the growth of the accumulated magnetic flux will proceed unimpeded. This latter point was emphasized by \cite{CK98}~(see their Figure~1b) and was missed by \cite{B-KLB02}. Real accretion disks may favor both types of field evolution at different stages of their lifetimes, namely outward flux diffusion followed by the generation of poloidal magnetic field loops around the disk's inner edge, and inward flux advection followed by field reconnection with the flux accumulated inside the inner edge of the disk \cite{C19}.

The field growth cannot proceed beyond equipartition at the inner edge of the disk, because the innermost accretion flow will be severely disrupted, and our analysis will break down. We can estimate a timescale $\tau_{\rm CB}$ for the innermost magnetic field to grow to equipartition if we assume that the system radiates continuously at roughly its Eddington value. Dimensional analysis of eq.~(\ref{inductionintegrated}) yields
\begin{equation}\label{tauCB}
\tau_{\rm CB}\sim \frac{eB_{\rm eq}r_{\rm ISCO}}{cf_{\rm rad}^\phi}\sim \frac{4\pi e^2B_{\rm eq}r^3_{\rm ISCO}}{L\sigma_{\rm T}}\sim \frac{eB_{\rm eq}r^3_{\rm ISCO}}{cGMm_p}\ .
\end{equation}
In the case of a stellar-mass black hole with $B_{\rm eq}\sim 10^7$~G, this characteristic time is on the order of a few weeks. $\tau_{\rm CB}$ scales roughly as $M^{3/2}$ with black hole mass, and therefore, the corresponding time for supermassive black holes is on the order of a few million years. We emphasize that these estimates were obtained for accretion flows that radiate continuously at close to their Eddington limit. In realistic astrophysical sources such as X-ray binary outbursts and AGNs, this is obviously not the case. For a more detailed review, the reader may consult \cite{C15}.

In summary, the Cosmic Battery is a special case of the well-known Biermann Battery according to which any force on the plasma electrons that is misaligned with respect to the direction of the gradient of the plasma density generates a non-zero $\nabla\times {\bf E}$ \cite{B50}. Radiation pressure is in general such a force, only {\em we have considered the astrophysical environment where it is maximal and maximally misaligned, namely the inner edge of the accretion disk around an astrophysical black hole radiating at Eddington luminosity levels}.
The Biermann Battery is in general very inefficient, and generates only a seed field that requires further dynamo amplification. The aberration of radiation due to rotation has been considered in the past in relation to the growth of the solar magnetic field, but was also found to be extremely inefficient \cite{CS66}. On the contrary, {\em the Cosmic Battery is the Biermann Battery `on steroids'}. It does not need extra dynamo amplification since it can steadily grow the magnetic field at the inner edge of the accretion disk from zero to astrophysically significant values within a few hundred million dynamical times. 

\section{Confirmation of the Cosmic Battery}

\subsection{Numerical simulations}

Over the years, we performed several numerical simulations of the Cosmic Battery in astrophysical accretion disks. Our first simulations were simple 1D integrations of eq.~(\ref{induction})  \cite{CK98}. We considered then a simple Shakura-Sunyaev disk \cite{SS73} with constant ratio of the local flow temperature to the virial temperature and constant ratio of magnetic diffusivity to viscosity (the inverse of the Prandtl number). We obtained the then unexpected result discussed in the previous section that, when diffusion dominates, the return polarity of the poloidal magnetic field loops generated around the inner edge of the disk diffuses continuously outward through the disk. This is what allows the magnetic flux to grow linearly with time around the central black hole. As we discussed already, this steady growth cannot continue beyond the point that the field becomes so strong that it dissrupts the innermost accretion flow, i.e. the growth is limited by equipartition. On the contrary, when inward field advection dominates over outward diffusion, the growth of the accumulated field quickly saturates over a few accretion dynamical times, and the Cosmic Battery does not generate astrophysically significant magnetic fields. This delicate balance (or imbalance) between flux advection and diffusion characterized by the magnetic Prandtl number is, we believe, central in understanding the dynamics of X-ray binary outbursts, and even the radio loud/radio quiet dichotomy \cite{Ketal12, C19}.
\begin{figure}[t]
\centering
\vspace{-2cm}
\includegraphics[width=15 cm]{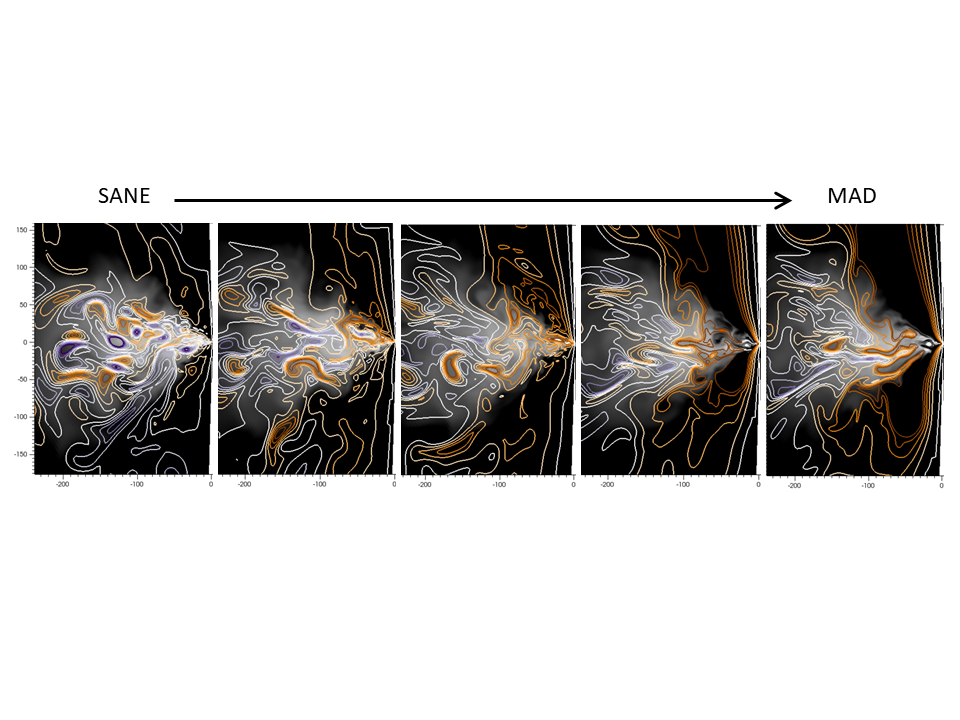}
\vspace{-3.5cm}
\caption{Schematic evolution of an ideal 2.5D GR-RMHD numerical simulation of an accretion flow (cloudy wedge region) from a SANE to a MAD state after the Cosmic Battery term is activated in the induction equation (adapted from \cite{Cetal18}). Lengths in units of the gravitational radius of the central Schwarzschild black hole. Poloidal magnetic field lines: orange: counter-clockwise, purple: clockwise. The disk accretes due to the MRI. Initially, the magnetic field is random. With time (from left to right), the poloidal magnetic field grows steadily as predicted by the Cosmic Battery (${\bf B}$ parallel to $\Omega$ around the center, ${\bf B}$ antiparallel to $\Omega$ in the bulk of the disk). The return magnetic field escapes outward through the disk due to numerical diffusivity. The simulation is stopped when the magnetic field accumulated around the center becomes so strong that the innermost accretion flow is arrested.}
\vspace{-0.3cm}
\label{CosmicBattery2}
\end{figure}

We performed several more numerical simulations in 2.5D 					\cite{CKC06, CCK08, CNK15}, mainly to refute the criticism of \cite{LRB-K09}, but didn't learn much more than what we knew since the original 1D simulations of \cite{CK98}. In all cases, the radiation field was treated as coming from a central point source. In reality, however, the radiation field originates in the accretion-ejection flow itself. Our first effort to consider the complex radiation field in the immediate vicinity of the central black hole led to the 2.5D general relativistic radiation magneto-hydrodynamic (GR-RMHD) numerical simulations of \cite{Cetal18}. The initial configuration was a thick torus threaded by a weak poloidal magnetic field of alternating polarity. We did not want to impose a preferred polarity, and let the system accrete on its own via the Magneto-Rotational Instability (MRI \cite{BH00}). Indeed, the system reached a so-called Standard And Normal Evolution (SANE) with steady accretion and no preferred magnetic field polarity. Initially, the Cosmic Battery term in the induction equation was switched off. After the system reached the SANE steady-state, we continued the original simulation, but also started a new simulation with the Cosmic Battery term turned on. We then saw clearly what was expected since \cite{CK98}: poloidal magnetic field loops were generated, their interior field polarity (${\bf B} || \Omega$) steadily accumulated around the central black hole, while their return field polarity diffused continuously outwards through the accretion flow. Notice that our simulations were performed under ideal MHD conditions, and yet numerical diffusion introduced the effect required to see the Cosmic Battery in action. Thus, we were in some sense lucky. We interrupted our simulation when the innermost accretion flow was arrested by the magnetic stresses due to the field accumulated around the central black hole. More numerical work is needed to understand other aspects of radiation in accretion-ejection flows around astrophysical black holes (e.g. the role of the optical depth).

\subsection{Observations}

We believe that there is great unexplored potential in establishing the action of some kind of Battery in astrophysical systems with observations of {\em magnetic field asymmetries}. As is well known, MHD involves only quadratic terms in the magnetic field through the Lorentz and electric forces $J\times B\sim (\nabla \times B)\times B\sim B^2$ and $\rho_e E\sim (\nabla\cdot E)E\sim E^2\sim B^2$ respectively. This implies that if we stop a numerical MHD simulation, reverse all magnetic fields, and then re-start the simulation, the evolution will be indistinguishable from the uninterrupted one. In other words, the direction of the magnetic field is not expected to depend on the direction of the flow, unless, there is some kind of Battery mechanism.

Lynden-Bell explored magnetic asymmetries in a comparison between the sign of the Faraday RM in the central regions of spiral galaxies seen edge on, and the directon of galactic rotation \cite{L13}. The sign of the Faraday RM is determined by the direction of the line-of-sight magnetic field, and the direction of galactic rotation is determined by the geometry of the spirals. In 8 out of a total of 9 cases for which he was able to obtain RM data, he found that ${\bf B} || \Omega$ in the central parts of the galaxies! Such magnetic field asymmetry is in agreement with the prediction of the Cosmic Battery and cannot be explained with pure MHD processes.

Another asymmetry has to do with the helical structure of the magnetic field in the jet, and, in particular, with the direction of its axial electric current. It is straightforward to see that, if the direction of the poloidal magnetic field is one-to-one related to the direction of rotation in the disk, then the rotation will `wind' the field and generate an azimuthal magnetic field component in a certain direction. If one then considers an astrophysical flow rotating in the opposite direction, the resulting poloidal magnetic field will be directed in the opposite direction. The rotation will then wind the field in the opposite direction resulting in {\em the same} direction of the azimuthal magnetic field. Indeed, the Cosmic Battery predicts that the directions of axial electric currents in the jet are universal. The axial electric current in the electron-positron black hole jet flows {\em towards the black hole} and returns along the current sheet along the last open magnetic field lines that crosses the black hole horizon. The axial electric current in the electron-proton disk wind/jet flows {\em away from the black hole} and returns at larger cylindrical radii outside the jet (see figure~2 of \cite{Cetal16} for details).

A definite direction of axial electric current is related to a steady Faraday RM gradient accross the jet, wherever Faraday RMs are observable. Unfortunately, observing such a gradient turned out to be so difficult that many researchers even doubt that the large scale helical magnetic field geometry required by MHD models of astrophysical jets is real (or at least they argue that it is not observable). Nevertheless, out of a total of about 89 cases of kpc-scale radio jets with Faraday RMs, we were able to observe steady gradients accross parts of the jet in only 18 cases\footnote{These include the large scale magnetic field in the two hemispheres of our own Galaxy which, technically, does not contain a jet.} \cite{Cetal09, Ketal12, Cetal16}. In all cases, the direction of the axial electric current is outwards!  This is another example of statistically significant asymmetry in agreement with the prediction of the Cosmic Battery that cannot be explained with pure MHD processes.

In conclusion, we argue that the Cosmic Battery enters as a previously unaccounted for player in our search for the origin of cosmic magnetic fields and in our investigation of accretion-ejection flows around astrophysical black holes. Ongoing observations of magnetic field asymmetries may one day establish the Cosmic Battery as a fundamental physical mechanism in astrophysics.

\end{document}